**Topology induced anomalous plasmon modes in metallic Möbius nanorings**

*Yin Yin*[1,2,*], *Shilong Li*[1], *Vivienne Engemaier*[1], *Ehsan Saei Ghareh Naz*[1], *Silvia Giudicatti*[1], *Libo Ma*[1,*], and *Oliver G. Schmidt*[1,2]

[*] Corresponding author: Y.Y. (y.yin@ifw-dresden.de); L.B.M. (l.ma@ifw-dresden.de)

[1]Institute for Integrative Nanosciences, IFW Dresden, Helmholtzstr. 20, 01069 Dresden, Germany

[2]Material Systems for Nanoelectronics, Technische Universität Chemnitz, Reichenhainer Str. 70, 09107 Chemnitz, Germany

**Abstract:**

We report on the investigation of plasmonic resonances in metallic Möbius nanorings. Half-integer numbers of resonant modes are observed due to the presence of an extra phase π provided by the topology of the Möbius nanostrip. Anomalous plasmon modes located at the non-orientable surface of the Möbius nanoring break the symmetry that exist in conventional ring cavities, thus enable far-field excitation and emission as bright modes. The far-field resonant wavelength as well as the feature of half-integer mode numbers is invariant to the change of charge distribution on the Möbius nanoring due to the nontrivial topology. Owing to the ultra-small mode volume induced by the remaining dark feature, an extremely high sensitivity as well as a remarkable figure of merit is obtained in sensing performance. The topological metallic nanostructure provides a novel platform for the investigation of localized surface plasmon modes exhibiting unique phenomena in plasmonic applications such as high sensitive detection and plasmonic nanolasers.

## 1. Introduction

Metallic nanostructures, capable of supporting surface plasmon polaritons (SPPs) and localized surface plasmon resonances, sit at the heart of subwavelength nanophotonic devices which can overcome traditional diffraction limits.[1-9] Among others, plasmonic nanoresonators such as nanorings are expected to be essential elements of future subwavelength-scale photonic systems.[10-12] The optical response and the plasmon modes of a metallic nanoresonator critically depend on its shape and size. Dipole-like modes, which have been widely investigated in metallic nanostructures,[13,14] represent the basic resonance form of localized surface plasmons. In particular, metallic nanoring resonators are known to support whispering-gallery-like plasmonic multiple modes, formed by self-interference of SPPs in the ring structures. Owing to the constraint of resonance, only integer numbers of plasmonic modes exist in nanoring resonators.[15,16] Multiple plasmon modes with antisymmetric charge distributions are known to exhibit inhibited radiative losses (so-called dark plasmon modes) due to the absence of any net dipole moment. However, dark plasmon modes can be converted into bright ones (i.e. with efficient radiative emission) by breaking the symmetry of the plasmon mode distributions, but so far have received little attention in plasmonic ring resonators.[17-21]

In this work, we report on the occurrence of topology-induced half-integer plasmon modes in metallic Möbius nanorings. Due to symmetry breaking, the higher-order plasmon modes turn into bright ones in the Möbius configuration, which are supposed to be dark in conventional cylindrical nanorings. The feature of half-integer numbers of plasmon modes as well as the corresponding resonant frequencies is robust to the variation of the surface-charge distribution on the Möbius nanoring due to the non-trivial topology. In addition, the bright plasmon modes retain some "dark" features, such as an enhanced quality (Q) factor and intense near-field confinement, leading to high sensitivity to refractive-index (RI) fluctuations together with a remarkable figure of merit (FOM).

## 2. Modeling and simulation

The Möbius nanoring is formed by rolling up a silver nanostrip after performing a half-twist, as schematically illustrated in **Figure 1**(a). The dimension of the nanostrip is set as $w = 80$ nm in width, $t = 10$ nm in thickness, and $L = 300\pi$ nm in length. For comparison, a cylindrical ring is also constructed by rolling up the same nanostrip without twist. The Drude model is adopted to describe the dispersive permittivity of silver as $\varepsilon(\omega) = \varepsilon_\infty - \omega_p^2/(\omega^2 + i\omega\gamma)$, where the high-frequency bulk permittivity is $\varepsilon_\infty = 6$, the plasma frequency is $\omega_p = 1.5 \times 10^{16}$ rad/s and the collision frequency is $\gamma = 7.73 \times 10^{13}$ rad/s.[22] The surrounding medium is air with a refractive index of $n = 1$. Linearly polarized light is employed to excite the plasmon modes in the nanoring structures. All the calculations are performed based on the finite-element method using the commercial software COMSOL.

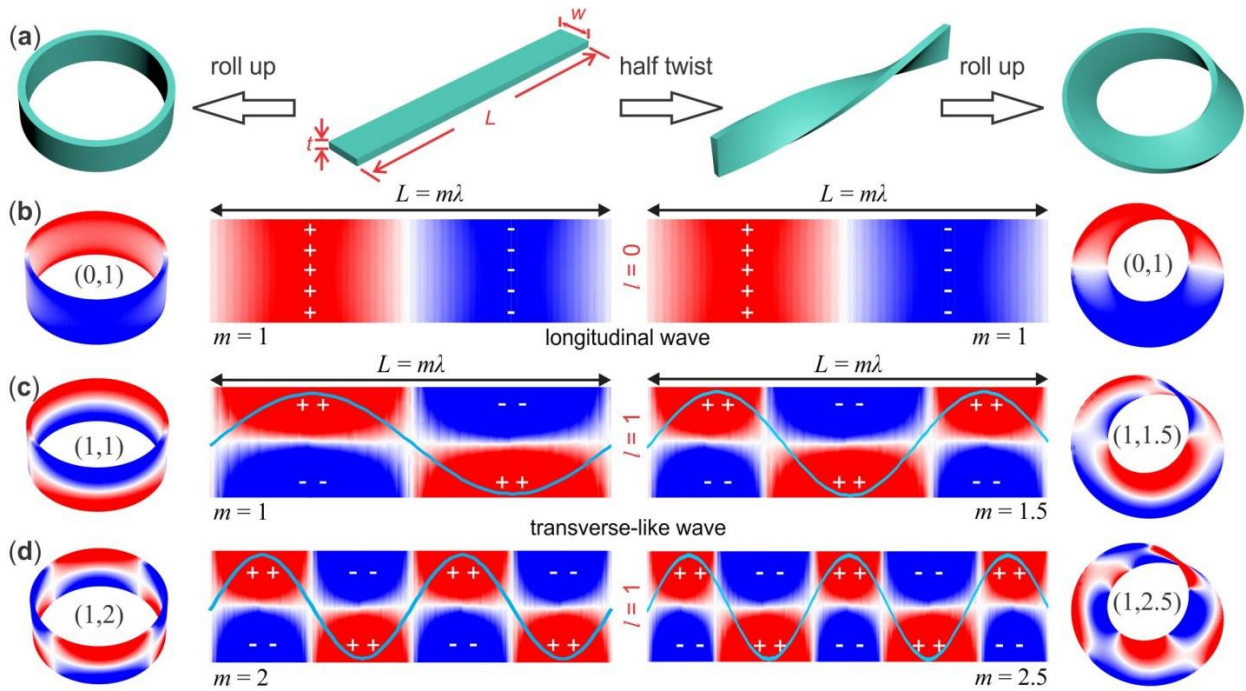

Figure 1. (a) A cylindrical nanoring made by rolling up a planar nanostrip (left panel), and a Möbius nanoring formed by rolling up the nanostrip after performing a half-twist (right panel). (b) Similar dipole-like plasmonic resonant modes are identified in both cylindrical and Möbius nanorings, formed by longitudinal wave oscillation in the ring structures. (c) and (d) show distinct antisymmetric higher order plasmon modes in the cylindrical and Möbius ring, formed by interferences of integer and half-integer number of transverse-like waves, respectively, as indicated by the solid sinusoids. The charge distributions of the plasmon modes are represented by false color, shown in both the rings and the corresponding planar strip forms.

## 3. Results and discussions

## 3.1 Plasmon modes in cylindrical and Möbius nanorings

The localized plasmonic resonances can be viewed as quasistatic SPPs confined within the nanostructure.[23] When SPPs propagate along the ring trajectory in the cylindrical and Möbius nanostrip, resonances are established when satisfying the resonant condition, i.e. an integer number ($m$) of wavelength λ fits into the perimeter (L) of the trajectory. In addition, for a higher-order resonance SPPs can also oscillate along the lateral direction of the strip with an integer number ($l$) of the plasmon mode. In this sense, each plasmon mode in the nanoring can be identified by a ($l$, $m$) pair. As the basic plasmonic resonance, the dipole plasmon-modes (0, 1) are found in both cylindrical and Möbius nanorings, as shown in Figure 1(b). Similar to previous reports, the dipole plasmon-mode is formed by charge oscillations as a longitudinal wave in the metallic ring structure.[15,16]

For the higher-order plasmonic resonances, the SPPs start to oscillate along the lateral direction of the nanostrip in addition to oscillating in the azimuthal plane of the ring cavities. The lateral oscillation can be viewed as a transverse-like wave with respect to the ring trajectories, where one pair of plasmonic antinodes constitutes a full wavelength (see Figure 1(c) and (d)). For the formation of constructive resonances, the resonant waves are required to be in-phase, or equivalently a phase difference of an integer number of 2π after one-round trip along the ring trajectory needs to be present to satisfy the resonant condition. In the cylindrical nanoring, integer numbers (e.g. $m$=1, 2) of waves are identified, and each wavelength carries a phase change of 2π, satisfying the resonant condition. This phenomenon is well known in optical WGM resonances in ring cavities.[24] In contrast to the cylindrical nanoring, half-integer numbers (i.e. $m$=1.5, 2.5) of oscillation waves for resonances are supported in the metallic Möbius nanoring. The occurrence of half-integer waves leads to a phase π missing for satisfying the resonant condition, thus contradicts the conventional picture of constructive interference. The formation of half-integer waves, i.e. half-integer plasmon mode pairs, is explained by the occurrence of an extra phase (also called Berry phase) π provided by the Möbius nanostrip, which has been reported in Möbius-ring resonators for both radio and visible frequency resonances, as well as in the half-twisted band structure of topological insulators for spin transports.[25-28]

## 3.2 Far-field spectra of the cylindrical and Möbius nanorings

*3.2.1 Bright modes occurred in Möbius nanorings*

The surface charge distribution of anomalous plasmon modes on the non-orientable surface of a Möbius nanoring breaks the symmetry that exists in conventional cylindrical rings, consequently leading to unique near and far field properties. Intrinsic plasmon modes possessing a net dipole moment, such as the dipole-like plasmon modes, are referred to as "bright modes", which can be directly excited and readily characterized by far-field techniques. On the contrary, higher order plasmon modes having vanishing net dipole moments are called "dark modes", which cannot be excited nor detected in the far-field due to efficient inhibition of radiative losses. To investigate the far-field properties, transmission (*T*) and reflectance (*R*) spectra of a cylindrical ring and a Möbius ring are calculated, as shown in **Figure 2**(a) and (b), respectively. Only one resonant peak located at 1408 nm is recognized in the cylindrical ring, which is induced by the charge oscillation of the dipole mode (0, 1), as shown in the right inset of Figure 2(a). The charge distribution in the dipole mode possesses $C_{1v}$ symmetry, manifesting a net dipole moment. In contrast, for the higher order plasmon modes shown in Figure 1, the charge oscillations exhibit antisymmetric distributions with $D_{1h}$ and $D_{2h}$ symmetries for modes (1, 1) and (1, 2), respectively. Hence, these modes do not occur in either the transmission or reflectance spectrum due to the absence of a net dipole moment, as has been reported in circular plasmonic ring resonators.[15,16,20,29] The near-field calculation results of (1, 1) and (1, 2) plasmon modes show that their optical resonant wavelengths are located at 565, and 539 nm, respectively, as indicated by the dotted line in Figure 2(a).

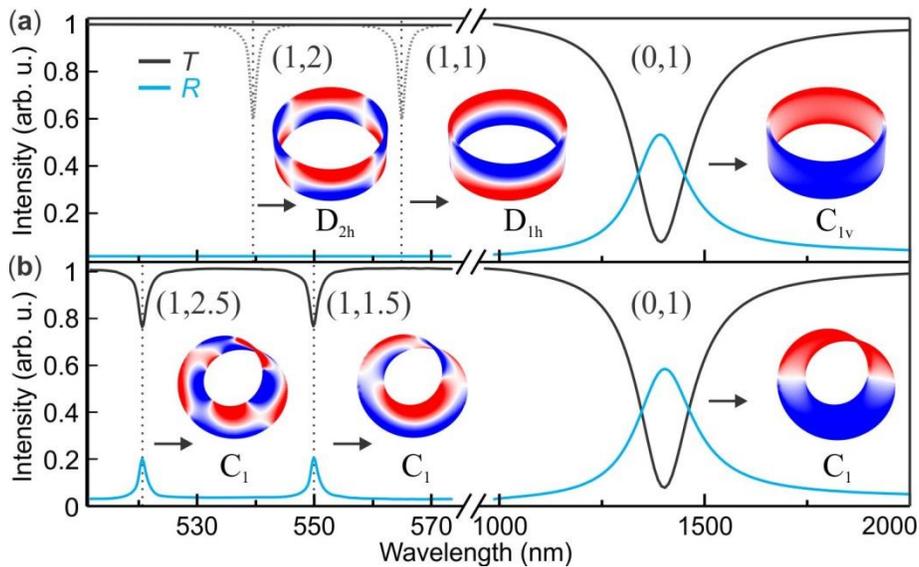

Figure 2. Transmission (*T*) and reflectance (*R*) spectra of a cylindrical ring (a) and Möbius ring (b). Bright modes (0, 1) caused by dipole-like surface charge oscillation are observed in both cylindrical and Möbius rings. The higher order plasmon modes exhibit dark feature in the cylindrical ring due to the antisymmetric surface charge distribution, while they turn to be bright in the Möbius ring due to the broken symmetry. The dark modes are indicated by dashed curves. The insets display the near-field antinodes distributions of corresponding modes in the ring cavities.

For the Möbius nanoring, the resonant peak of the dipole mode (0, 1) is located at 1416 nm, as shown in Figure 2(b). Although the length of the nanostrip in the Möbius ring is the same as that of the cylindrical ring, a slight mode shift of 8 nm is observed due to the geometry difference between the cylindrical and Möbius nanoring. In contrast to the cylindrical nanoring where the higher order plasmon modes exhibit dark features, efficient far-field emission is observed at 550 and 522 nm for (1, 1.5) and (1, 2.5) plasmonic modes, respectively, as shown in Figure 2(b). The intrinsic resonances with half-integer mode numbers violate the resonant condition in conventional rings,[25,26,30,31] which, however, can be supported by the topological Möbius ring. Here, the half-integer modes located at the non-orientable plane of the Möbius ring break the constraint of zero net dipole moment in conventional rings and enable the capability of far-field emission.

The Q-factor of dipole plasmon-modes is as low as in previous reports on circular plasmonic ring resonators owing to significant radiative losses.[15,16,20,29] Interestingly, the Q-factors of higher-order modes (1, 1.5) and (1, 2.5) are around six times higher than that of the dipole mode. The Q-factor enhancement indicates a smaller mode volume for the higher-order plasmon modes in the Möbius nanoring. Because of the small mode volume, the higher-order plasmon modes exhibit suppressed radiative loss, enhanced near-field intensities, and therefore higher Q-factors. In the Möbius nanoring, the mode volume is as small as $0.003\lambda_0^3$ calculated by the formula $V_{eff} = \int \varepsilon(r)E(r)^2 d^3r /(\varepsilon E_{max}^2)$, where $\lambda_0$ is the resonant wavelength, $\varepsilon(r)$ the permittivity and $E(r)$ is the electric field strength. The small mode volume indicates a strong local electric field with an enhancement factor of $(E_{max}/E_0)^2 \approx 1000$ ($E_{max}$ is the local field at the plasmon mode and $E_0$ is the electric field of excitation wave), which is promising for the investigation of light-matter interactions.

*3.2.2 Symmetry-breaking in Möbius nanorings*

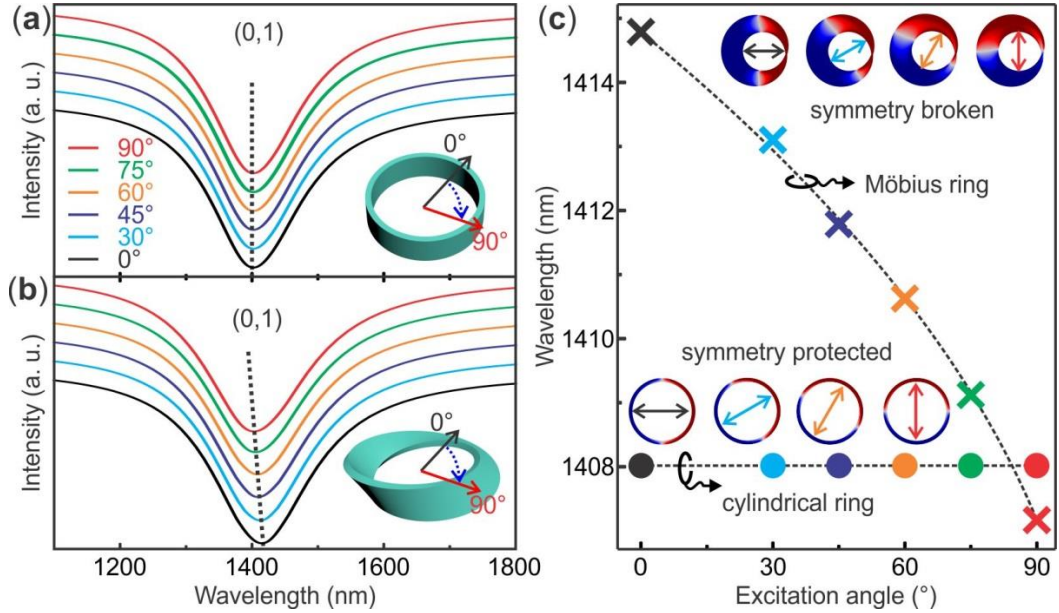

Figure 3. Transmission spectra of the dipole mode (0, 1) in cylindrical (a) and Möbius rings (b) excited under different polarization angles. (c) Upon varying the excitation polarization, the resonant wavelength is constant in cylindrical nanoring due to the rotational symmetry and shifts significantly in Möbius nanoring owing to the broken symmetry. The insets show representative surface charge distributions in the two nanoring structures.

The far-field resonant peaks as a function of the excitation orientation are shown in **Figure 3**, in which the plane wave with different polarization orientations in the azimuthal plane is used to excite the plasmon modes in the cylindrical and Möbius ring. Cylindrical rings possess rotational symmetry with respect to the ring axis (*z*). Therefore, the resonant wavelength and surface charge distribution of both dipole and high order plasmon modes are invariant to the change of excitation-polarization orientation. As an example, the transmission spectra of dipole modes are shown in Figure 3(a), showing the constant resonant peak when being excited under different polarization angles (see Figure 3(c)). On the contrary, the Möbius ring has no such rotational symmetry. Thus, the surface-charge distribution varies when the polarization orientation of the excitation light changes, as shown in the insets of Figure 3(c). As a consequence, the far-field resonant peaks of the dipole mode do vary when excited at different polarization orientations, as shown in Figure 3(a).

*3.2.3 Invariant resonant wavelength in Möbius nanorings*

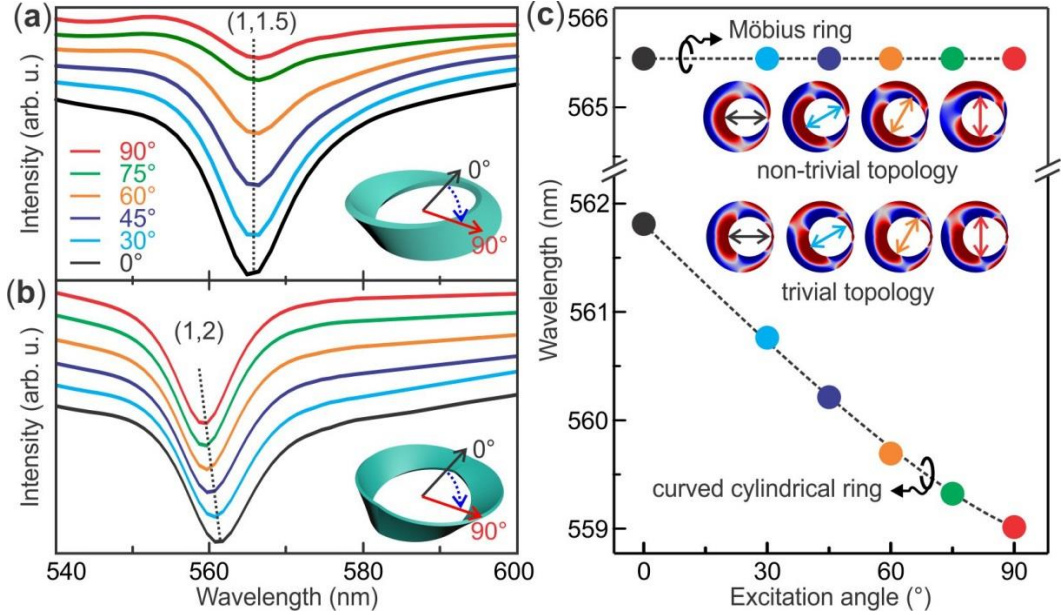

Figure 4. Transmission spectra of the high order mode (1, 1.5) in Möbius (a) and mode (1,2) in curved cylindrical ring (b) excited with different polarization angles. (c) Upon varying the excitation polarization, the resonant wavelength is constant in Möbius nanoring due to the topological structure, while shifts significantly in curved cylindrical nanoring of trivial topology. The insets show the representative surface charge distributions in the two nanoring structures.

Interestingly, for the topology induced half-integer modes such as (1, 1.5), the resonant wavelength is constant despite the changes of the charge distribution, as shown in **Figure 4**(a) and (c). Compared with the dipole mode shown in Figure 3(c) which occurs independently of topology, the half-integer mode is protected by the topology of the Möbius nanoring. The invariance of the resonant wavelength indicates that the net dipole moment in total is constant no matter how the charge distribution of the half-integer plasmon mode changes on the topological Möbius ring. This topology origin can be further verified by the same examination in a cylindrical-like nanoring composed of a curved but not twisted nanostrip, i.e. having the same topology of the cylindrical ring. One can see that the higher order plasmon mode (1, 2) shows the integer number feature, and the corresponding resonant wavelength varies when excited with different polarization orientations, as shown in Figure 4(b) and (c). This difference indicates that the resonant peak invariance is induced by the topology of the Möbius ring, rather than the curvature of the nanostrip. Moreover, the mode intensity in the transmission spectrum varies when excited with different polarization orientations. This transmission variation is caused by the change of extinction cross section when exciting the Möbius ring with different polarization orientations.

### 3.3 High sensitivity of the bright mode in Möbius nanoring

As the higher order plasmon modes in the Möbius ring retain some dark feature, a high sensitivity is expected due to the enhanced local field. Here the mode (1, 1.5) was selected as an example to demonstrate the sensing performance. As shown in **Figure 5**, a significant mode shift is observed when slightly changing the RI of the surrounding medium down to $\Delta n = 0.005$. The sensitivity reaches 1000 nm per refractive index unit (RIU), exhibiting an excellent bright-mode-based sensing performance. A FOM of 100 is calculated when the sensitivity is divided by the resonance line width, which is remarkable for a plasmonic resonator.[13,14] In addition, the intrinsic absorption loss of metallic structures can be compensated by introducing active/optical gain media which can efficiently improve the Q-factor of plasmonic nanosystems.[32] Therefore, an even higher FOM can be expected, which is not only attractive for sensing applications but also interesting for nonlinear optics, such as plasmonic nanolasers.[5,7,9]

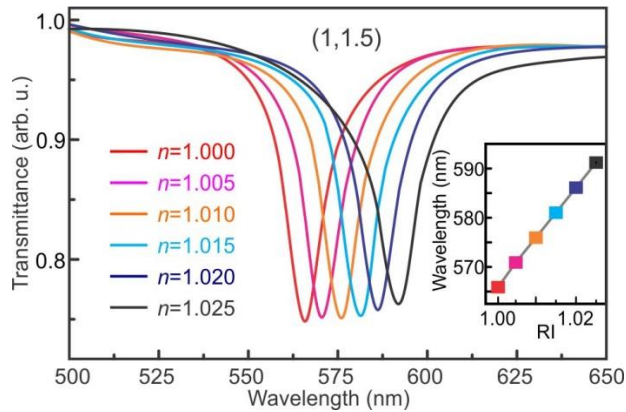

Figure 5. Resonance shifts of the plasmon mode (1, 1.5) in response to RI change of the surrounding medium. The resonant wavelength as a function of the RI is shown in the inset.

### 4. Conclusion

In conclusion, topological Möbius nanorings have been designed to study higher order plasmon modes. Anomalous half-integer numbers of resonant modes are observed which is caused by the presence of an extra phase π provided by the topology of the Möbius nanostrip. The distribution of the unusual plasmon modes at the non-orientable surface of Möbius nanoring breaks the symmetry that exists in conventional ring cavities, therefore exhibits bright-mode features for far-field excitation and emission. The half-integer plasmon modes together with the far-field emission wavelength are invariant to the mode distributions on the Möbius nanoring due to the non-trivial topology. Owing to the remaining dark feature, a high sensitivity of 1000 nm/RIU as well as a remarkable FOM of 100 is achieved for sensing applications extending from the near-field to the far-field. The topological metallic nanostructure provides a novel platform for the investigation of localized plasmon modes exhibiting unique phenomena in plasmonic applications such as high sensitive detection and plasmonic nanolasers.

## Acknowledgements

This work was supported by the DFG research group FOR 1713. Y.Y. acknowledges support by China Scholarship Council under file No. 201206090008.

## References:

[1]   S. A. Maier, *Plasmonics: fundamentals and applications* (Springer Science & Business Media, 2007).
[2]   S. Nie and S. R. Emory, Science **275**, 1102-1106 (1997).
[3]   T. W. Ebbesen, H. J. Lezec, H. Ghaemi, T. Thio, and P. Wolff, Nature **391**, 667-669 (1998).
[4]   W. L. Barnes, A. Dereux, and T. W. Ebbesen, Nature **424**, 824-830 (2003).
[5]   D. J. Bergman and M. I. Stockman, Phys. Rev. Lett. **90**, 027402 (2003).
[6]   S. I. Bozhevolnyi, V. S. Volkov, E. Devaux, J.-Y. Laluet, and T. W. Ebbesen, Nature **440**, 508-511 (2006).
[7]   R. F. Oulton, V. J. Sorger, T. Zentgraf, R.-M. Ma, C. Gladden, L. Dai, G. Bartal, and X. Zhang, Nature **461**, 629-632 (2009).
[8]   P. Berini and I. De Leon, Nat. Photonics **6**, 16-24 (2012).
[9]   Y. Yin, T. Qiu, J. Li, and P. K. Chu, Nano Energy **1**, 25-41 (2012).
[10]   E. M. Larsson, J. Alegret, M. Käll, and D. S. Sutherland, Nano Lett. **7**, 1256-1263 (2007).
[11]   A. W. Clark and J. M. Cooper, Small **7**, 119-125 (2011).
[12]   C.-Y. Tsai, J.-W. Lin, C.-Y. Wu, P.-T. Lin, T.-W. Lu, and P.-T. Lee, Nano Lett. **12**, 1648-1654 (2012).
[13]   K. A. Willets and R. P. Van Duyne, Annu. Rev. Phys. Chem. **58**, 267-297 (2007).
[14]   K. M. Mayer and J. H. Hafner, Chem. Rev. **111**, 3828-3857 (2011).
[15]   J. Aizpurua, P. Hanarp, D. Sutherland, M. Käll, G. W. Bryant, and F. G. De Abajo, Phys. Rev. Lett. **90**, 057401 (2003).

[16]   F. Hao, E. M. Larsson, T. A. Ali, D. S. Sutherland, and P. Nordlander, Chem. Phys. Lett. **458**, 262-266 (2008).
[17]   S. Zhang, D. A. Genov, Y. Wang, M. Liu, and X. Zhang, Phys. Rev. Lett. **101**, 047401 (2008).
[18]   N. Liu, L. Langguth, T. Weiss, J. Kästel, M. Fleischhauer, T. Pfau, and H. Giessen, Nat. Mater. **8**, 758-762 (2009).
[19]   B. Luk'yanchuk, N. I. Zheludev, S. A. Maier, N. J. Halas, P. Nordlander, H. Giessen, and C. T. Chong, Nat. Mater. **9**, 707-715 (2010).
[20]   F. Hao, Y. Sonnefraud, P. V. Dorpe, S. A. Maier, N. J. Halas, and P. Nordlander, Nano Lett. **8**, 3983-3988 (2008).
[21]   D. Gómez, Z. Teo, M. Altissimo, T. Davis, S. Earl, and A. Roberts, Nano Lett. **13**, 3722-3728 (2013).
[22]   Y. Liu, T. Zentgraf, G. Bartal, and X. Zhang, Nano Lett. **10**, 1991-1997 (2010).
[23]   J. Le Perchec, P. Quemerais, A. Barbara, and T. Lopez-Rios, Phys. Rev. Lett. **100**, 066408 (2008).
[24]   K. J. Vahala, Nature **424**, 839-846 (2003).
[25]   D. J. Ballon and H. U. Voss, Phys. Rev. Lett. **101**, 247701 (2008).
[26]   S. Li, L. Ma, V. Fomin, S. Böttner, M. Jorgensen, and O. G. Schmidt, arXiv preprint arXiv:1311.7158 (2013).
[27]   H. C. Manoharan, Nat. Nanotechnology **5**, 477-479 (2010).
[28]   L. Ma, S. Li, V. Fomin, M. Hentschel, J. Götte, Y. Yin, M. Jorgensen, and O. G. Schmidt, Nature communications **7** (2016).
[29]   E. J. R. Vesseur and A. Polman, Nano Lett. **11**, 5524-5530 (2011).
[30]   E. Miliordos, Phys. Rev. A **82**, 062118 (2010).
[31]   Z. Li and L. Ram-Mohan, Phys. Rev. B **85**, 195438 (2012).
[32]   Z.-G. Dong, H. Liu, J.-X. Cao, T. Li, S.-M. Wang, S.-N. Zhu, and X. Zhang, Appl. Phys. Lett. **97**, 114101 (2010).